# Interactive Physics-Inspired Traffic Congestion Management

Hossein Rastgoftar

*Abstract*—This paper proposes a new physics-based approach to effectively control congestion in a network of interconnected roads (NOIR). The paper integrates mass flow conservation and diffusion-based dynamics to model traffic coordination in a NOIR. The mass conservation law is used to model the traffic density dynamics across the NOIR while the diffusion law is applied to include traffic speed and motion direction into planning. This paper offers an analogy between traffic coordination in a transportation system and heat flux in a thermal system to define a potential filed over the NOIR. The paper also develops an interactive light-based and boundary control to manage traffic congestion through optimizing the traffic signal operations and controlling traffic flows at the NOIR boundary nodes. More specifically, a model predictive boundary control optimizes the NOIR inflow traffic while a receding horizon optimizer assigns the optimal movement phases at the NOIR intersections. For simulation, the paper models traffic congestion in a heterogeneous NOIR with a large number of interior and boundary nodes where the proposed interactive control can successfully manage the congestion.

## I. INTRODUCTION

TRAFFIC is a complex phenomenon dealing with coordination of many vehicles in a network of interconnected roads (NOIR). While the number of vehicles in an NOIR can change with time, the space contained by the NOIR is fixed and limited. Traffic congestion is an important issue that wastes time and fuel and can cause accident in highways. Therefore, efficient traffic congestion management models are highly demanded as they can positively impact several aspects of human life as well as the environment. This paper will offer a new interactive physic-based traffic control (IPBTC) approach to model and control traffic congestion in a heterogeneous NOIR consisting of unidirectional and bidirectional roads. The proposed IPBTC can address the urban mobility key challenges by improving the traffic speed which in turn results in reducing car emissions.

Researchers have suggested a variety of approaches to manage traffic congestion. The traditional fixed-cycle control [1], [2] has been applied by the traffic signals to manage vehicle coordination at intersections. A fixed cycle control is commonly specified by the cycle time, split time, and offset time to characterize cycle completion, green light duration, and delays, respectively. Traffic network study tool (TRANSYT) [3], [4] is the existing software used for optimizing the signal timing under the fixed-cycle assumption.

More recently, the traffic responsive control methods have been proposed by the researchers due to advancement in sensing, computing, and vision algorithms. The available approaches can be classified as model-free and model-based methods. Model-free methods suggested in [5]–[12] use the reinforcement learning (RL) to adaptively mange the traffic signal operations and traffic coordination in a transportation infrastructure. Model-based traffic control approaches commonly use the conservation law and model the traffic dynamics by a linear system [13]–[16]. In particular, cell transition model (CTM) [17]–[19] has been used to fill every road by a finite number of serially-connected road elements and develop a first-order conservation-based traffic dynamics by discretizing the mass flow conservation PDE. Model-based reinforcement learning using Markov decision process offered in [20]–[22] to manage traffic congestion. Particularly, single agent RL [5] and multi-agent RL [8] have been previously proposed to optimize the traffic signal movement phases. Refs. [23]– [26] apply MPC to assign optimal traffic speed in highways. MPC have been also applied in Refs. [27]–[29] to optimize the traffic signal operations. Additionally, researchers have suggested fuzzy logic [30]–[33], neural network [34]–[37], mixed nonlinear programming (MNLP) [38] approaches for the model-based traffic management. Furthermore, traffic congestion management is modeled as a boundary control problem in Refs. [39], [40]. In Refs. [39], [40], a first-order traffic dynamics inspired by mass flow conservation is developed to the traffic coordination freeways.

This paper aims to advance the resilience and efficiency of the traffic congestion control by incorporating the macroscopic and microscopic traffic behavior into planning. To this end, the paper effectively integrates conduction-based boundary control and conservation-based light control to manage the traffic congestion. Particularly, an interactive physics-based traffic control (IPBTC) will be developed to optimize (i) discrete actions commanded by the traffic lights at the NOIR intersections, and (ii) traffic inflow rates at the NOIR boundary

H. Rastgoftar is with the Department of Aerospace Engineering, University of Michigan, Ann Arbor, MI, 48109 USA e-mail: hosseinr@umich.edu.

nodes. IPBTC models the traffic macroscopic coordination as a conduction problem governed by a parabolic PDE with spatio-temporal parameters. Conduction-based coordination is inspired by the following two facts:

1) Traffic must move at a certain direction assigned by an authorized decision-maker (ADM) at every NOIR (See Figure 1).
2) The traffic speed is a distributed non-negative quantity assuming vehicles obey the prescribed motion direction at any location of the NOIR.

These two features of traffic coordination are analogous to conduction in a thermal system. More specifically, heat flux is the gradient of the temperature distribution in a thermal system where it is always conducted from high-temperature locations toward low-temperature locations. Similarly, the paper defines a scalar potential field over the NOIR which is analogous to temperature distribution in a thermal system. By defining the traffic speed as the gradient of the traffic potential function, traffic flows from a high-potential location to a low-potential location. Furthermore, IPBTC applies the mass conservation law to model the microscopic traffic congestion. Particularly, the microscopic traffic coordination is modeled by a stochastic switching dynamics obtained by governing the mass conservation law at every road element across the NOIR.

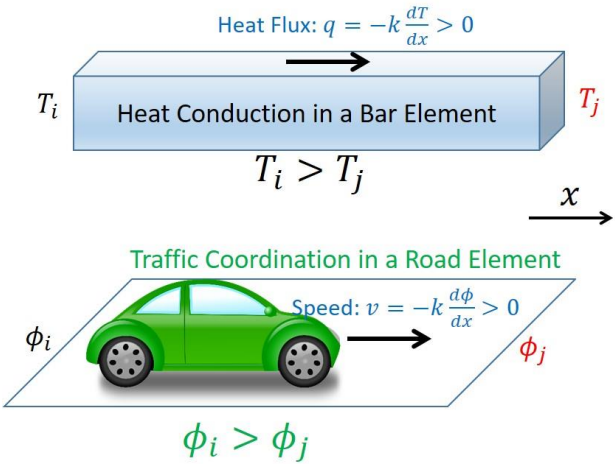

Figure 1: Analogy between traffic coordination and heat diffusion. Traffic coordinates from a high-potential location $i$ (with potential value $\phi_i$) toward a low-potential location $j$ (with potential value $\phi_j$) where traffic speed is always positive at every location across the road element. Similarly, heat flux is directed from a high-temperature location toward a low-temperature location in a thermal system.

Compared to the existing literature, this papers offers the following contributions:

1) This paper offers a novel interactive light-based and boundary congestion control approach.

2) A new probabilistic traffic dynamics inspired by mass flow conservation will be developed where both traffic inflow and outflow rates are included to obtain the traffic dynamics at every road element (This will be discussed more in Remark 2).

3) The proposed probabilistic dynamics ensures that the traffic density is positive at any time t. Therefore, the MPC boundary controller can control the traffic congestion with minimum computation cost only by imposing the inequality constraints on the boundary traffic inflow rates without imposing state constraints.

4) An interface between conduction-based and conservation dynamics will be defined in the papers. Therefore, parameters of the conduction-based dynamics are uniquely related to the parameters of the mass conservation-based dynamics.

This paper is organized as follows: Preliminary notions on graph theory and discrete state transition are presented in Section II. Problem statement in Section III is followed by the proposed physics-based traffic coordination model given in Section IV. Interactive boundary and light-based traffic control model is developed in Section V. Simulation results presented in Section VI are followed by concluding remarks in Section VII.

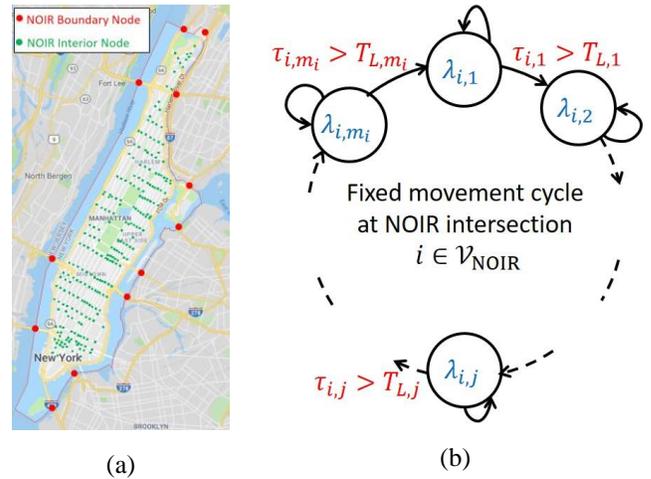

Figure 2: (a) Example NOIR-Manhattan. NOIR boundary and interior nodes are shown by red and green, respectively[1]. (b) Fixed movement cycle at NOIR intersection $i \in \mathcal{V}_C$ modeled by a non-deterministic finite automata.

## II. PRELIMINARIES

This paper defines the following three graphs to model and control the traffic congestion:

1. **NOIR graph** $\mathcal{G}_{\text{NOIR}}(\mathcal{V}_{\text{NOIR}}, \mathcal{E}_{\text{NOIR}})$ *with node set* $\mathcal{V}_{\text{NOIR}} = \{1, \cdots, m\}$ defining all NOIR intersections and edge set $\mathcal{E}_{\text{NOIR}} \subset \mathcal{V}_{\text{NOIR}} \times \mathcal{V}_{\text{NOIR}}$. $\mathcal{V}_{\text{NOIR}} = \mathcal{V}_B \cup \mathcal{V}_C$ where $\mathcal{V}_B = \{1, \cdots, m_B\}$ and $\mathcal{V}_C = \{m_B + $

---

[1] The background image was downloaded from Google Map.

1, ⋯, m}. An example NOIR defined over Manhattan is shown in Figure 2.

2. **BC traffic network** $\mathcal{G}(\mathcal{V}, \mathcal{E})$ with node set $\mathcal{V}$ and edge set $\mathcal{E} \subset \mathcal{V} \times \mathcal{V}$. A node $i \in \mathcal{V}$ represent a road element with specified motion direction. Graph G is used to study the macroscopic traffic behavior.

3. **Switching LBRC traffic network** $\mathcal{G}_{\text{LBRC},\lambda}(\mathcal{V}, \mathcal{E}_\lambda)$ with node set $\mathcal{V}$ and edge set $\mathcal{E}_\lambda \subset \mathcal{V} \times \mathcal{V}$. Note that $\lambda \in \Lambda$ and $\mathcal{E}_\lambda \subset \mathcal{E}$, $\Lambda$ is a finite set defining traffic network topologies commanded by traffic lights at the NOIR intersections, i.e. $\Lambda$ determines every possible movement phase across the NOIR junctions.

Movement phases are defined by finite set

$$\Lambda = \lambda_{m_B+1} \times \cdots \times \lambda_m,$$

where $\lambda_i$ specifies all possible movement phases at intersection $i \in \mathcal{V}_C$, and "×" is the Cartesian product symbol. In particular, it is assumed that mi unique movements can be commanded at $i \in \mathcal{V}_{\text{NOIR}}$, therefore,

$$\lambda_i = \{\lambda_{i,1}, \cdots, \lambda_{m_i,1}\}, \quad i \in \mathcal{V}_C.$$

In this paper, we make the following assumptions:

1. A unique movement phase $\lambda_{i,j}[k]$ ($j = 1, \cdots, m_i$) can be commanded at node $i \in \mathcal{V}_C$ over the time interval $[t_{k-1}, t_k]$.
2. Transition of movement phases across the NOIR is defined by a non-deterministic finite state automaton [41] at every intersection node $i \in \mathcal{V}_C$ (See Figure 2 (b)).
3. An upper bound time limit $T_{L,i}$ is considered for the duration of movement $\lambda_{i,j}$ at intersection $i \in \mathcal{V}_C$. A local timer measures activation period $\tau_{i,j}$ of the movement phase $\lambda_{i,j}$ at NOIR intersection $i \in \mathcal{V}_C$. Once $\tau_{i,j} = T_{L,i}$, the movement phase $\lambda_{i,j}$ is overridden at $i \in \mathcal{V}_C$.

Given the time-limit threshold $T_{L,i}$ at every node $i \in \mathcal{V}_C$,

$$\Lambda^{next}[k] = \lambda^{next}_{m_B+1}[k] \times \cdots \times \lambda^{next}_m[k] \subset \Lambda$$

specifies next allowable movement phases at time $k + 1$ given movement phases at current time $k$. This paper assumes that the movement cycle is fixed at every NOIR intersection. Therefore, $|\lambda_i^{next}| \leq 2$ at any time $k$ ($\forall i \in \mathcal{V}_C$). $|\lambda_i^{next}| = 2$, if $\tau_{i,j} < T_{L,i}$ when $\lambda_{i,j}$ is active at NOIR intersection node $i \in \mathcal{V}_C$. Otherwise, the ADM overrides $\lambda_{i,j}$ and $|\lambda_i^{next}|$ drops to 1. Consequently, $\Lambda^{next} \leq 2^{m-m_B}$. Figure 2 (b) shows an example four-leg NOIR intersection node $i \in \mathcal{V}_C$ with eight possible movements.

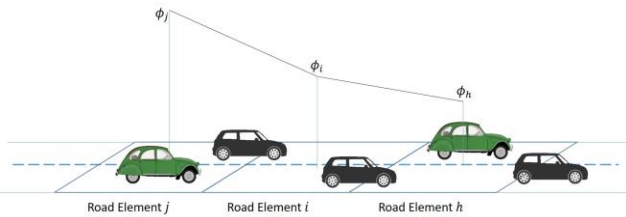

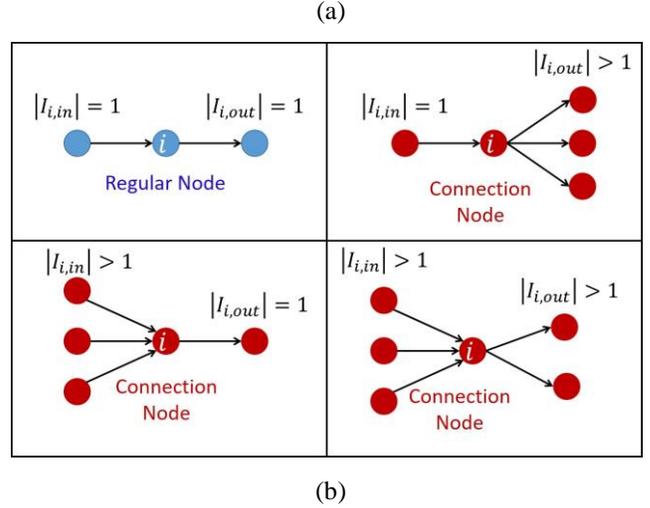

Figure 3: (a) Discretizing a road into a finite number of serially-connected bar elements. It is assumed that the traffic potential function $\phi$ is linearly decreasing across a road element. Therefore, $V_{i,out}$ and $V_{i,in}$ are positive quantities at every node $i \in \mathcal{V}$. (b) Schematic of regular and connection road elements.

## III. PROBLEM STATEMENT

This paper studies the problem of traffic congestion control in a large heterogeneous NOIR consisting of bidirectional and unidirectional roads. This problem is splitted into the following two problems:

1. **Traffic Boundary Control** through controlling the traffic inflows at the NOIR boundary nodes.
2. **Traffic Light-Based Responsive Control** (LBRC) through optimizing discrete actions commanded by the traffic lights situated at the NOIR intersection nodes.

For the traffic boundary control, traffic coordination is modeled as a diffusion problem governed by a parabolic PDE. A scalar potential field $\phi(\mathbf{r}, t)$ is defined over the NOIR where $\mathbf{r}$ and $t$ denote position and time, respectively. Potential function $\phi(\mathbf{r}, t)$ is similar to temperature distribution in a thermal system and governed by the following PDE:

$$C(\mathbf{r}, t)\, \frac{\partial \phi(\mathbf{r},t)}{\partial t} = \nabla \cdot (K(\mathbf{r}, t) \nabla \phi(\mathbf{r}, t)) \qquad (1)$$

where $C(\mathbf{r}, t)$ is the specific coordination capacity (SCC) and $K(\mathbf{r}, t)$ is the distributed stiffness (conductivity). The conduction-inspired macroscopic traffic coordination is managed through controlling traffic inflows at the NOIR boundary nodes. The paper defines a distributed traffic rate (speed) $V: \mathbb{R}^2 \times \mathbb{R} \to \mathbb{R}_+$ over the NOIR. $V(\mathbf{r}, t)$, specifying the number of vehicles passing from NOIR position $\mathbf{r}$ at time $t$, has the following properties:

1. It is a positive quantity everywhere across the NOIR at any time $t$.
2. $V(\mathbf{r}, t)$ is related to $\phi(\mathbf{r}, t)$ by $V(\mathbf{r}, t) = -\nabla \phi(\mathbf{r}, t)$.

3. Because $V(\mathbf{r},t)$ is positive everywhere in the NOIR, $\phi(\mathbf{r},t)$ is decreasing, if we move along the motion direction in a NOIR road (See Figure 3 (a)).

To model the traffic coordination, the governing Eq. (1) is spatially discretized using the finite element method (FEM). The BC traffic network $\mathcal{G}(\mathcal{V},\mathcal{E})$ is defined over the NOIR, where $\mathcal{V} = \{1,\cdots,N\}$ is the node set and $\mathcal{E}\subset\mathcal{V}\times\mathcal{V}$ defines edges of $\mathcal{G}$. It is assumed that NOIR roads are filled by serially-connected road elements where $i\in\mathcal{V}$ represents a unique road element. The node set $\mathcal{V}$ can be expressed as

$$\mathcal{V} = \mathcal{V}_{in}\cup\mathcal{V}_{out}\cup\mathcal{V}_I$$

where $\mathcal{V}_{in} = \{1,\cdots,N_{in}\}$, $\mathcal{V}_{out} = \{N_{in}+1,\cdots,N_{out}\}$, and $\mathcal{V}_I = \{N_{out}+1,\cdots,N\}$ define index numbers of inlet, outlet, and interior road elements, respectively. Every node $i\in\mathcal{V}$ has a neighbor set $\mathcal{I}_i$ defined as follows:

$$\mathcal{I}_i = \{\mathcal{I}_{i,in},\mathcal{I}_{i,out}\}, \quad (2)$$

where $\mathcal{I}_{i,in}$ and $\mathcal{I}_{i,out}$ specify node i's in-neighbor and out-neighbor nodes. Traffic enters $i\in\mathcal{V}$ from an in-neighbor node belonging to $\mathcal{I}_{i,in}$ and it exits from $i\in\mathcal{V}$ towards an out-neighbor node belonging to $\mathcal{I}_{i,out}$. $i\in\mathcal{V}$ is a regular node, if $|\mathcal{I}_{i,in}| = |\mathcal{I}_{i,out}| = 1$. Otherwise, $i\in\mathcal{V}$ is a connection node. Schematic of regular and connection road elements are shown in Figure 3 (b).

For the traffic LBRC, the microscopic traffic coordination is treated as a mass conservation problem governed by the continuity PDE. The switching LBRC graph $\mathcal{G}_{LBRC,\lambda}(\mathcal{V},\mathcal{E}_\lambda)$ is used to spatially discretize continuity PDE when $\lambda\in\Lambda$ defines the traffic network topology. Consequently, the conservationbased traffic dynamics is modeled by a stochastic linear system and the Receding Horizon Optimization (RHO) [42] is used to optimize the discrete actions (movement phases) commanded by the traffic lights.

## IV. TRAFFIC COORDINATION MODEL

A conduction-inspired traffic model will be developed in Section IV-A and used to control the traffic coordination through the NOIR boundary nodes. Section IV-B models traffic coordination by a discrete-time dynamics inspired by the mass conservation law. The relation between parameters of conduction-based and conservation-based dynamics is also discussed in Section IV-B.

### A. Macroscopic Traffic Coordination as a Conduction Problem

Consider the schematic of road element $i\in\mathcal{V}$ shown in Fig. 4. Every road element i has a unique specific coordination capacity (SCC) $c_i > 0$. Let $\phi_i$ be the potential value of the element $i$, $\tilde{V}_{i,in} > 0$ denote the **weighted inflow traffic rate**, and $\tilde{V}_{i,out} > 0$ denote the **weighted outflow traffic rate**. The inflow and outflow traffic rates are defined as the number of vehicles entering and leaving the bar element in a certain amount of time (See Figure 4). It is assumed that the conduction-based traffic coordination satisfies the following four rules:

**Rule 1:** $c_i \dfrac{d\phi_i}{dt} = \tilde{V}_{i,in} - \tilde{V}_{i,out}$

**Rule 2:** $\tilde{V}_{i,in} = \begin{cases}\sum_{j\in\mathcal{I}_{i,in}}\kappa_{i,j}(\phi_j(t)-\phi_i(t)) & i\in\mathcal{V}\setminus\mathcal{V}_{in}\\ \text{given} & i\in\mathcal{V}_{in}\end{cases}$ (3)

**Rule 3:** $\tilde{V}_{i,out} = \begin{cases}\sum_{i\in\mathcal{I}_{i,out}}\kappa_{i,h}(\phi_i(t)-\phi_h(t)) & i\in\mathcal{V}\setminus\mathcal{V}_{out}\\ \text{given} & i\in\mathcal{V}_{out}\end{cases}$

**Rule 4:** $\phi_i(t) = 0, \quad \forall t, \forall i\in\mathcal{V}_{out}$,

where $\kappa_{i,j} > 0$ is the stiffness between road elements $i\in\mathcal{V}$ and $j\in\mathcal{I}_{i,in}$. Rule 4 implies that the potential values of those road elements defined by $\mathcal{V}_{out}$ are all zero.

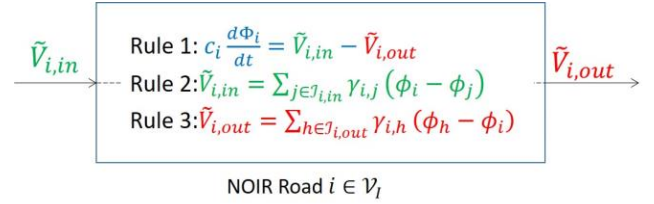

(a)

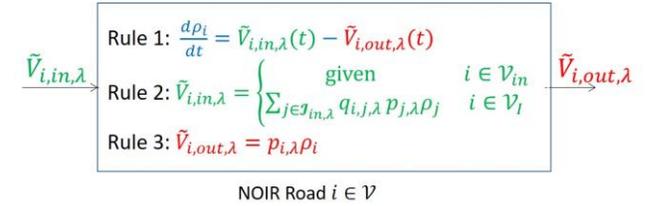

(b)

Figure 4: (a) Example road element $i\in\mathcal{V}$. Rules 1, 2, and 3 for obtaining the conduction-inspired traffic coordination. (b) Traffic flow inspired by mass conservation law.

**Relation Between non-weighted Outflow and Potential Value:** Let $\mathbf{\Phi}(t) = [\mathbf{\Phi}_{in}^T\ \mathbf{\Phi}_I^T]^T\in\mathbb{R}^{(N-N_{out}+N_{in})\times 1}$ define nodal potential values, where $\mathbf{\Phi}_{in} = [\phi_1\ \cdots\ \phi_{N_{in}}]^T$ and $\mathbf{\Phi}_I = [\phi_{N_{out}+1}\ \cdots\ \phi_N]^T$. Defining non-weighted traffic outflow inlet

$$V_{i,out} = \begin{cases}\sum_{h\in\mathcal{I}_{i,out}}(\phi_i(t)-\phi_h(t)) & i\in\mathcal{V}\setminus\mathcal{V}_{out}\\ \text{given} & i\in\mathcal{V}_{out}\end{cases},$$

$\mathbf{V}(t) = [V_{1,out}\ \cdots\ V_{N_{in},out}\ V_{N_{out}+1,out}\ \cdots\ V_{N,out}]$ and $\mathbf{\Phi}(t)$ can be related by

$$\mathbf{V}(t) = \mathbf{L}\mathbf{\Phi}(t). \quad (4)$$

This paper assumes that $V_{i,out}$ measurement is available at every road element $i\in\mathcal{V}$ at all times $t$. Defining $\zeta_i = i+N_{out}-N_{in}$, $\mathbf{L} = [L_{ij}]\in\mathbb{R}^{(N-N_{out}+N_{in})\times(N-N_{out}+N_{in})}$ is given by

$$L_{ij} = \begin{cases}\sum_{j\in\mathcal{I}_{i,out}}1 & (i\in\mathcal{V}_{in}\wedge j=i)\vee(\zeta_i\in\mathcal{V}_I\wedge j=i)\\ -1 & i\in\mathcal{V}_{in}\wedge j=\mathcal{I}_{i,out}\\ -1 & \zeta_i\in\mathcal{V}_I\wedge\zeta_j\in\mathcal{I}_{\zeta_i,out}\setminus(\mathcal{I}_{\zeta_j,out}\cap\mathcal{V}_{out})\\ 0 & \text{otherwise.}\end{cases} \quad (5)$$

It is assumed that every inlet road element is connected to a single interior node, therefore, $L_{ij} = |\mathcal{I}_{i,out}| = 1$, if $i \in \mathcal{V}_{in}$.

*Proposition 1:* Matrix **L** is nonsingular with the eigenvalues that are all located on the right-hand side of the $s$-plane.

*Proof:* Because **L** is defined by Eq. (5), it is diagonally dominant. Diagonal entries of **L** are all positive, therefore, **L** can be converted to $\tilde{\mathbf{L}} = \mathbf{\Gamma L}$ where $\mathbf{\Gamma} = [\Gamma_{ij}] \in \mathbb{R}^{(N-N_{out}+N_{in})\times(N-N_{out}+N_{in})}$ is positive definite and diagonal, and $\Gamma_{ii} = \frac{1}{L_{ii}}$ ($i = 1, \cdots, N + N_{out} - N_{in}$). While sum of the row elements is 0 in $N - N_{out} - N_{in}$ rows of $\tilde{\mathbf{L}}$, the remaining rows of $\tilde{\mathbf{L}}$ are negative sum rows. Therefore, Furthermore, $\tilde{\mathbf{L}} = \mathbf{I} - \mathbf{D}$ is non-singular M-matrix where spectral radius of matrix **D** is less than 1. Consequently, eigenvalues of $\tilde{\mathbf{L}}$ are all placed on the right-hand side of the $s$-plane which in turn implies that the real part of all eigenvalues of matrix **L** are positive.

*Proposition 2:* Matrix $\mathbf{L}^{-1}$ is non-negative.

*Proof:* While diagonal entries of matrix L are all positive, off-diagonal entries of **L** are either 0 or negative. Using the Gauss-Jordan elimination method, the augmented matrix $\mathbf{L}_a = [\mathbf{L} \;\vdots\; \mathbf{I}] \in \mathbb{R}^{(N-N_{out}+N_{in})\times 2(N-N_{out}+N_{in})}$ can be converted to matrix $\tilde{\mathbf{L}}_a = [\mathbf{I} \;\vdots\; \mathbf{L}^{-1}] \in \mathbb{R}^{(N-N_{out}+N_{in})\times 2(N-N_{out}+N_{in})}$ only by performing row algebraic operations. Every entry of the lower triangle of matrix **L** can be converted to 0 if a top row is multiplied by a positive scalar and the outcome is added to the other rows. Elements of the upper triangle of matrix **L** can be similarly vanished, if a bottom row is multiplied by a positive scalar and the outcome is added to the other rows. Therefore, **L**$^{-1}$, obtained by performing these row operations on **L**, is non-negative.

**Remark 1:** Because $\mathbf{L}^{-1}$ is non-negative and nonsingular and $\mathbf{V}(t) \geq \mathbf{0}$ at any time $t$, nodal potential vector $\dot{\mathbf{\Phi}}$ is nonnegative which in turn implies that $\phi_i(t) \geq 0$ at any node $i \in \mathcal{V}_I \cup \mathcal{V}_{in}$ at all times $t$.

Traffic Network Dynamics: Assuming SCC parameter $c_i$ is known at every road element $i \in \mathcal{V}$, the SCC matrix

$$\mathbf{C} = \text{diag}(c_1, \cdots, c_{N_{in}}, c_{N_{out}+1}, \cdots c_N)$$

is positive-definite and diagonal. Furthermore, it is assumed that stiffness $\kappa_{i,j} > 0$ is known for every element $i \in \mathcal{V}$ with in-neighbor element $j \in \mathcal{I}_i$; stiffness matrix $\mathbf{K} = [K_{ij}] \in \mathbb{R}^{N \times N}$ is assigned as follows:

$$K_{ij} = \begin{cases} \kappa_{i,j} & j \in \mathcal{I}_i \wedge i \in \mathcal{V} \\ -\sum_{j \in \mathcal{I}_i} \kappa_{i,j} & j = i \wedge i \in \mathcal{V} \\ 0 & \text{otherwise.} \end{cases} \quad (6)$$

Matrix **K** is negative semi-definite. The first eigenvalue of **K** is 0 while the remaining **K** eigenvalues are all negative. Let

$$\mathbf{K} = \begin{bmatrix} \mathbf{K}_{in,in} & \mathbf{K}_{in,out} & \mathbf{K}_{in,I} \\ \mathbf{K}_{out,in} & \mathbf{K}_{out,out} & \mathbf{K}_{out,I} \\ \mathbf{K}_{I,in} & \mathbf{K}_{I,out} & \mathbf{K}_{I,I} \end{bmatrix}, \mathbf{K}_R = \begin{bmatrix} \mathbf{K}_{in,in} & \mathbf{K}_{in,I} \\ \mathbf{K}_{I,in} & \mathbf{K}_{I,I} \end{bmatrix}.$$

Then, $\mathbf{K}_R$ is negative definite. Define positive input vector $\mathbf{u}(t) = [V_{1,out} \cdots V_{N_{in},out}]^T$, and $\mathbf{W} = \begin{bmatrix} \mathbf{I} \\ \mathbf{0} \end{bmatrix} \in \mathbb{R}^{(N-N_{out}+N_{in}) \times N_{in}}$. Assuming matrices **C** and $\mathbf{K}_R$ are piecewise constant, the network traffic coordination dynamics is given by

$$\begin{cases} \dot{\mathbf{\Phi}} = \mathbf{A}\mathbf{\Phi} + \mathbf{B}\mathbf{u} \\ \mathbf{V} = \mathbf{L}\mathbf{\Phi} \end{cases}, \quad (7)$$

where $\mathbf{A} = \mathbf{C}^{-1}\mathbf{K}_R$ and $\mathbf{B} = \mathbf{C}^{-1}\mathbf{W}$.

*B. Microscopic Traffic Coordination as a Mass Conservation Problem*

Let $\rho_i(t)$ denote the number of vehicles in road element $i \in \mathcal{V}$ at time $t$. The switching LBRC network is defined by graph $\mathcal{G}_{\text{LBRC},\lambda}(\mathcal{V}, \mathcal{E}_\lambda)$, where $\lambda \in \Lambda$, and $\Lambda$ is a finite set defining all possible movement phases commanded by traffic lights at NOIR intersections defined by $\mathcal{V}_C = \{m_B + 1, \cdots, m\} \subset \mathcal{V}_{\text{NOIR}}$. As shown in Figure 4 (b), a road element $i \in \mathcal{V}$ can be considered as a control volume. $\bar{V}_{i,in,\lambda}$ and $\bar{V}_{i,out,\lambda}$ are the average traffic inflow and outflow, if $\lambda \in \Lambda$ specifies the topology of the traffic network. To develop the traffic coordination dynamics, it is assumed that the traffic flow obeys the following rules:

**Rule 1:** Traffic coordination satisfies the conservation law at every road element $i \in \mathcal{V}$. Therefore

$$i \in \mathcal{V}, \lambda \in \Lambda, \quad \frac{d\rho_i}{dt} = \bar{V}_{i,in,\lambda}(t) - \bar{V}_{i,out,\lambda}(t). \quad (8)$$

**Rule 2:** $p_{i,\lambda}$ is the fraction of the existing vesicles leaving NOIR road element $i \in \mathcal{V}$ at time $t$ when $\lambda \in \Lambda$ specifies the movement phases at time $t$. Therefore,

$$i \in \mathcal{V}, \lambda \in \Lambda, \quad \bar{V}_{i,out,\lambda}(t) = p_{i,\lambda}\rho_i(t). \quad (9)$$

**Rule 3:** Let $j \in \mathcal{I}_{i,out,\lambda}$ be an out-neighbor node of the road $i \in \mathcal{V} \setminus \mathcal{V}_{out}$ and $q_{j,i,\lambda}$ be the fraction of traffic flow entering $j \in \mathcal{I}_{i,out,\lambda}$ from $i \in \mathcal{V} \setminus \mathcal{V}_{out}$ when $\lambda \in \Lambda$ is active. Then,

$$\sum_{j \in \mathcal{I}_{i,out}} q_{j,i,\lambda} = 1$$

and the traffic inflow rate is given by

$$\begin{aligned} &i \in \mathcal{V}_{in}, \lambda \in \Lambda, \quad \bar{V}_{i,in,\lambda}(t) = \text{given}, \quad &(a) \\ &i \in \mathcal{V} \setminus \mathcal{V}_{in}, \lambda \in \Lambda, \quad \bar{V}_{i,in,\lambda}(t) = \quad &(10) \\ &\sum_{j \in \mathcal{I}_{i,in,\lambda}} q_{i,j,\lambda} \bar{V}_{j,out,\lambda}(t) = \sum_{j \in \mathcal{I}_{i,in,\lambda}} q_{i,j,\lambda} p_{j,\lambda} \rho_j \quad &(b) \end{aligned}$$

at road element $i \in \mathcal{V}$. Note that $\sum_{j \in \mathcal{I}_{i,in,\lambda}} q_{i,j,\lambda} > 0$ but it is not necessarily equal to 1. By substituting (9), (10) into (8), the traffic coordination dynamics is given by the following first-order dynamics:

$$\frac{d\rho_i}{dt} = \begin{cases} \bar{V}_{i,in,\lambda} - p_{i,\lambda}\rho_i & i \in \mathcal{V}_{in} \\ \sum_{j \in \mathcal{I}_{i,in,\lambda}} q_{i,j,\lambda} p_{j,\lambda} \rho_j - p_{i,\lambda}\rho_i & i \in \mathcal{V} \setminus \mathcal{V}_{in} \end{cases}. \quad (11)$$

for $\lambda \in \Lambda$.

*Theorem 1:* Assume $\bar{V}_{i,in,\lambda} > 0$ at every road element $i \in \mathcal{V}_{in}$, $\rho_i$ is updated by dynamics (11), $p_{i,\lambda} \geq 0$ at every road element

$i \in \mathcal{V}$, and $\rho_i(t_0) \geq 0$ at every road element $i \in \mathcal{V}$ at initial time $t_0$. Then, $\rho_i(t) \geq 0$ at any time $t \geq t_0$.

*Proof:* The number of cars at road element $i \in \mathcal{V}$ can become negative, if there exists a time $t_{n,i} \geq t_0$ such that (i) $\rho_i(t_{n,i}) = 0$, (ii) $\dot{\rho}_i(t_{n,i}^-) > 0$, and (iii) $\dot{\rho}_i(t_{n,i}^+) < 0$. **This situation never happens if $\rho_i$ is updated by dynamics** (11). If $p_{i,\lambda} = 0$, $\dot{\rho}_i(t) > 0$ at any time $t \geq t_0$ which in turn implies that $\rho_i(t) \geq 0$ assuming $\rho_i(t_0) \geq 0$ at every road element $i \in \mathcal{V}$. If $p_{i,\lambda} > 0$ and $\rho_i(t_0) \geq 0$, $\bar{V}_{i,out,\lambda}(t) = p_{i,\lambda}\rho_i(t)$, defined by (9), vanishes, only if $\rho_i(t_{n,i}^-) = 0$. Because $\rho_j(t_{n,i}^-) \geq 0$ for every other road element $j \in \mathcal{V}$ ($j \neq i$), $\bar{V}_{i,in,\lambda}(t_{n,i}) \geq 0$ for every road element $i \in \mathcal{V}$. Therefore, $\dot{\rho}_i(t_{n,i}^+) = \bar{V}_{i,in,\lambda}(t_{n,i}^+) - \bar{V}_{i,out,\lambda}(t_{n,i}^+) = \bar{V}_{i,in,\lambda}(t_{n,i}^+) \geq 0$ which in turn implies that $\rho_i(t) \geq 0$ at any time $t$.

**Remark 2:** A first-order traffic dynamic, inspired by mass flow conservation was previously developed in [39], [40]. In Refs. [39], [40], the traffic density is updated by

$$\frac{d\rho_i}{dt} = \sum_{j \in \mathcal{I}_{i,in}} q_{i,j}\rho_j - \rho_i$$

where $\sum_{j \in \mathcal{I}_{i,in}} q_{i,j} = 1$. This model does not include the traffic outflow rate to update the number of vehicles in road element (cell) $i$ at a time $t$. The proposed IPBTC applies new rules to determine the conservation-based traffic dynamics which in turn yields the new traffic dynamics (11) with a different structure. Note that Eq. (11) includes both traffic inflow and outflow rates for modeling of the traffic coordination at every road element $i \in \mathcal{V}$.

Defining $\rho(t) = [\rho_1 \cdots \rho_N]^T$, the LBRC dynamics is given by

$$\lambda \in \Lambda, \quad \frac{d\rho}{dt} = (-\mathbf{I} + \mathbf{P}_\lambda)\rho + \mathbf{W}\mathbf{u}. \tag{12}$$

where $\mathbf{W}$ and $\mathbf{u}$ were previously introduced in Section IV.A and $\mathbf{P}_\lambda$ and be partitioned as follows:

$$\lambda \in \Lambda, \quad \mathbf{P}_\lambda = \begin{bmatrix} \mathbf{P}_{in,in}^\lambda & \mathbf{P}_{in,out}^\lambda & \mathbf{P}_{in,I}^\lambda \\ \mathbf{P}_{out,in}^\lambda & \mathbf{P}_{out,out}^\lambda & \mathbf{P}_{out,I}^\lambda \\ \mathbf{P}_{I,in}^\lambda & \mathbf{P}_{I,out}^\lambda & \mathbf{P}_{I,I}^\lambda \end{bmatrix}.$$

Then, partition $\mathbf{P}_{\mathcal{X},\mathcal{Y}}^\lambda$ of $\mathbf{P}_\lambda$ specifies the fraction of traffic flow moving from $\mathcal{V}_\mathcal{Y}$ towards $\mathcal{V}_\mathcal{X}$, where $\mathcal{X}, \mathcal{Y} \in \{"in","out","I"\}$ are linguistic subscripts. For example, $\mathbf{P}_{in,out}^\lambda$ ($\mathcal{X} =$ "in" and $\mathcal{Y} =$ "out") specifies the percentage of traffic flow from $\mathcal{V}_{out}$ toward $\mathcal{V}_{in}$. $\mathbf{P}_\lambda = [P_{ij}^\lambda] \in \mathbb{R}^{N \times N}$ has the following properties:

1. It is non-negative, i.e. every element of matrix $\mathbf{P}_\lambda$ is non-negative.
2. Diagonal entry $P_{ii}^\lambda = 1 - p_{i,\lambda}$ specifies the percentage of vehicles remaining at road element $i \in \mathcal{V}$. Because $p_{i,\lambda} \in [0,1]$, $P_{ii}^\lambda = 1 - p_{i,\lambda} \in [0,1]$, for $i = 1,\cdots,N$.
3. Off-diagonal entry

$$i \in \mathcal{V} \setminus \mathcal{V}_{out}, j \neq i, \quad P_{ij}^\lambda = \begin{cases} q_{j,i,\lambda}p_{j,\lambda} & j \in \mathcal{I}_{i,out,\lambda} \\ 0 & j \notin \mathcal{I}_{i,out,\lambda} \end{cases}$$

specifies the percentage of the vehicles at node $i \in \mathcal{V} \setminus \mathcal{V}_{out}$ moving towards $j \in \mathcal{I}_{i,out,\lambda}$.

4. If $p_{i,\lambda} = 1$ for every "inlet" road element $i \in \mathcal{V}_{in}$, then, $\mathbf{P}_{in,in}^\lambda = \mathbf{0} \in \mathbb{R}^{N_{in} \times N_{in}}$. **This assumption implies that no vehicles are stored at the inlet road elements.** Similarly, $\mathbf{P}_{out,out}^\lambda = \mathbf{0} \in \mathbb{R}^{N_{out} \times N_{out}}$, if $p_{i,\lambda} = 1$ for every outlet node $i \in \mathcal{V}_{out}$. The latter assumption implies that no vehicles are stored at the outlet road elements.
5. $\mathbf{P}_{in,I}^\lambda = \mathbf{0} \in \mathbb{R}^{N_{in} \times (N-N_{out}+N_{in})}$ because no vehicle moves from $i \in \mathcal{V}_I$ towards $i \in \mathcal{V}_{in}$, i.e. $\mathcal{I}_{i,in,\lambda} = \emptyset$, if $i \in \mathcal{V}_{in}$.
6. $\mathbf{P}_{I,out}^\lambda = \mathbf{0} \in \mathbb{R}^{(N-N_{out}+N_{in}) \times N_{out}}$ and $\mathbf{P}_{in,out}^\lambda = \mathbf{0} \in \mathbb{R}^{N_{in} \times N_{out}}$ because no vehicle moves for $i \in \mathcal{V}_{out}$ toward $j \in \mathcal{V}_{in} \cup \mathcal{V}_I$, i.e. $\mathcal{I}_{i,out,\lambda} = \emptyset$, if $i \in \mathcal{V}_{out}$.

By admitting the above properties and assumptions, matrix $\mathbf{P}_\lambda$ simplifies to

$$\lambda \in \Lambda, \quad \mathbf{P}_\lambda = \begin{bmatrix} \mathbf{P}_{in,in}^\lambda & \mathbf{0} & \mathbf{0} \\ \mathbf{0} & \mathbf{0} & \mathbf{P}_{out,I}^\lambda \\ \mathbf{P}_{I,in}^\lambda & \mathbf{0} & \mathbf{P}_{I,I}^\lambda \end{bmatrix}.$$

*Proposition 3:* Sum of the entries of column $i \in \mathcal{V}_{out}$ of $\mathbf{P}_\lambda$ is zero but sum of the entries is 1 in the remaining column of matrix $\mathbf{P}_\lambda$.

*Proof:* As aforementioned, $\sum_{j \in \mathcal{I}_{i,out,\lambda}} q_{j,i,\lambda} = 1$ ($\forall \lambda \in \Lambda$), if $i \in \mathcal{V}_{in} \cup \mathcal{V}_I$. Sum of the elements of every column $i \in \{1,\cdots,N_{in},N_{out}+1,\cdots,N\}$ of matrix $\mathbf{P}_\lambda$ is 1 because

$$\sum_{j=1}^N P_{ij}^\lambda = (1 - p_{i,\lambda}) + p_{i,\lambda} \sum_{j \in \mathcal{I}_{i,out,\lambda}} q_{j,i,\lambda} = 1.$$

Defining

$$\mathbf{X}(t) = [\rho_1 \cdots \rho_{N_{in}} \rho_{N_{out}+1} \cdots \rho_N]^T,$$

the traffic dynamics (12) simplifies to

$$\dot{\mathbf{X}} = \mathbf{Q}_\lambda \mathbf{X} + \mathbf{W}\mathbf{u}, \tag{13}$$

where $\mathbf{W}$ and $\mathbf{u}$ were previously introduced in Section IV-A. $\mathbf{Q}_\lambda = -\mathbf{I} + \mathbf{P}_r^\lambda$ where

$$\mathbf{P}_r^\lambda = \begin{bmatrix} \mathbf{P}_{in,in}^\lambda & \mathbf{0} \\ \mathbf{P}_{I,in}^\lambda & \mathbf{P}_{I,I}^\lambda \end{bmatrix}.$$

Matrix $\mathbf{Q}_\lambda$ has the following properties:

1. Columns of $\mathbf{Q}_\lambda$ either sum to 0 or sum to a negative number.
2. Column $j$ of $\mathbf{Q}_\lambda$ sums to a negative number, if $j \in \mathcal{I}_{i,in,\lambda}$ and $i \in \mathcal{V}_{out}$. Otherwise, column $j$ sums to 0.

*Theorem 2:* If graph $\mathcal{G}_{LBRC,\lambda}(\mathcal{V},\mathcal{E}_\lambda)$ contains a path from every boundary node $i \in \mathcal{V}_{in}$ to each interior node $j \in \mathcal{V}_I$, then, matrix $\mathbf{Q}_\lambda$ is Hurwitz.

*Proof:* Because there exists a path from each boundary node to every interior node of graph $\mathcal{G}_{LBRC,\lambda}$, matrix $\mathbf{P}_r^\lambda$ is irreducible. Sum of the column entries is negative in at least one column of $\mathbf{Q}_\lambda$, the spectral radius of $\mathbf{P}_r^\lambda$, denoted by $r_\lambda$, is less than 1. Therefore, eigenvalues of $\mathbf{Q}_\lambda$ are all located inside a disk with radius $r_\lambda < 1$ with a center positioned at $-1 + 0j$ and $\mathbf{Q}_\lambda$ is Hurwitz [43], [44].

**Remark 3:** The outflow traffic rate (speed) vector

$$\mathbf{V}(t) = \begin{bmatrix} V_{1,out} & \cdots & V_{N_{in},out} & V_{N_{out}+1,out} & \cdots & V_{N,out} \end{bmatrix}$$

can be related to $\mathbf{X}(t)$ by

$$\mathbf{V}(t) = \mathbf{D}\mathbf{X}(t), \tag{14}$$

where $\mathbf{D} = \text{diag}(\bar{p}_1, \cdots, \bar{p}_{N_{in}}, \bar{p}_{N_{out}+1}, \cdots \bar{p}_N)$ is a positive definite and diagonal matrix. Note that

$$\bar{p}_i = \frac{\sum_{\lambda \in \Lambda} p_{i,\lambda}}{|\Lambda|} > 0$$

for every road element $i \in \mathcal{V}_{in} \cup \mathcal{V}_I$. Defining

$$\bar{\mathbf{Q}} = \frac{\sum_{\lambda \in \Lambda} \mathbf{Q}_\lambda}{|\Lambda|}$$

the average conservation-based dynamics is given by

$$\dot{\mathbf{X}} = \bar{\mathbf{Q}}\mathbf{X} + \mathbf{W}\mathbf{u}, \tag{15}$$

*Theorem 3:* Let $\bar{p}_i$ be known at every road element $i \in \mathcal{V}_{in}$. Then, conservation dynamics matrices $\bar{\mathbf{Q}}$ is related to the conduction-based matrix $\mathbf{A}$ using

$$\mathbf{A} = \mathbf{L}^{-1}\mathbf{D}\bar{\mathbf{Q}}\mathbf{D}^{-1}\mathbf{L}. \tag{16}$$

Furthermore, SCC parameter $c_i$ and $\bar{p}_i$ are related by

$$i \in \mathcal{V}_{in}, \quad c_i = \frac{1}{\bar{p}_i} \tag{17}$$

*Proof:* Considering Eq. (4), $\boldsymbol{\Phi}$ can be expressed as $\boldsymbol{\Phi} = \mathbf{L}^{-1}\mathbf{V}$ where $\mathbf{V} = \mathbf{D}\mathbf{X}$ (See Remark 3). Substituting $\boldsymbol{\Phi} = \mathbf{L}^{-1}\mathbf{D}\mathbf{X}$ into dynamics (7) yields

$$\dot{\mathbf{X}} = \underbrace{(\mathbf{D}^{-1}\mathbf{L}\mathbf{C}^{-1}\mathbf{K}_R\mathbf{L}^{-1}\mathbf{D})}_{\bar{\mathbf{Q}}}\mathbf{X} + \underbrace{(\mathbf{D}^{-1}\mathbf{L}\mathbf{C}^{-1}\mathbf{W})}_{\mathbf{W}}\mathbf{u} \tag{18}$$

Consequently, $\mathbf{A}$ can be uniquely specified based on the traffic tendency probabilities defined in Section IV-B. Let $\mathbf{W} = \begin{bmatrix} \mathbf{I} \\ \mathbf{0} \end{bmatrix}$, $\mathbf{L} = \begin{bmatrix} \mathbf{L}_{in,in} & \mathbf{L}_{in,I} \\ \mathbf{L}_{I,in} & \mathbf{L}_{I,I} \end{bmatrix}$, $\mathbf{D} = \begin{bmatrix} \mathbf{D}_{in,in} & \mathbf{0} \\ \mathbf{0} & \mathbf{D}_{I,I} \end{bmatrix}$, and $\mathbf{C} = \begin{bmatrix} \mathbf{C}_{in,in} & \mathbf{0} \\ \mathbf{0} & \mathbf{C}_{I,I} \end{bmatrix}$, $\mathbf{D}_{in,in} = \text{diag}(\bar{p}_1, \cdots, \bar{p}_{N_{in}})$, $\mathbf{D}_{I,I} = \text{diag}(\bar{p}_{N_{out}+1}, \cdots, \bar{p}_N)$, $\mathbf{C}_{in,in} = \text{diag}(c_1, \cdots, c_{N_{in}})$, and $\mathbf{C}_{I,I} = \text{diag}(c_{N_{out}+1}, \cdots, c_N)$ are positive definite and diagonal.

$$\mathbf{D}^{-1}\mathbf{L}\mathbf{C}^{-1}\mathbf{W} = \begin{bmatrix} \mathbf{D}_{in,in}^{-1}\mathbf{L}_{in,in}\mathbf{C}_{in,in}^{-1} \\ \mathbf{D}_{I,I}^{-1}\mathbf{L}_{I,in}\mathbf{C}_{in,in}^{-1} \end{bmatrix}.$$

Because traffic does not flow from any interior node toward an inlet node, $\mathbf{L}_{I,in} = \mathbf{0}$ which in turn implies that $\mathbf{D}_{I,I}^{-1}\mathbf{L}_{I,in}\mathbf{C}_{in,in}^{-1} = \mathbf{0}$. Therefore, $\mathbf{D}_{in,in}^{-1}\mathbf{L}_{in,in}\mathbf{C}_{in,in}^{-1} = \mathbf{I}$. Because an inlet node is only connected to one interior node, $\mathbf{L}_{in,in} = \mathbf{I}$ and $\mathbf{I}$ and $c_i$ is related to $\bar{p}_i$ using relation (17) for every inlet node $i \in \mathcal{V}_{in}$.

## V. Traffic Congestion Control

This paper offers an interactive physics-inspired control to efficiently manage congestion through the NOIR boundary and intersection nodes. A MPC-based boundary control will be developed in Section V-A to manage the macroscopic traffic coordination through the NOIR inlet boundary nodes. Furthermore, Section V-B formulates a receding horizon optimization model for assigning optimal actions commanded by the traffic lights at the NOIR intersections.

### A. Macroscopic Boundary Control

Assuming the traffic coordination dynamics is given by (15), the following difference equation is used to model the macroscopic traffic dynamics

$$\mathbf{X}[k+1] = \bar{\mathbf{Q}}_D\mathbf{X}[k] + \bar{\mathbf{W}}_D\mathbf{u}[k],$$

where $\bar{\mathbf{Q}}_D = \mathbf{I} + \Delta T\bar{\mathbf{Q}}$, $\bar{\mathbf{W}}_D = \Delta T\mathbf{W}$, and time increment $\Delta T$ is fixed. The optimization cost is defined by

$$\text{Cost}[k] = \frac{1}{2}\sum_{\tau=1}^{n_\tau}(\mathbf{X}^T[k+\tau]\mathbf{X}[k+\tau] + \beta\mathbf{u}^T[k+\tau]\mathbf{u}[k+\tau]), \tag{19}$$

subject to equality constraint $\mathbf{u}[k] = u_0$ and $\mathbf{u}[k] > 0$ at any discrete time $k$, where $\beta > 0$ is a scaling factor and $n_t$ is constant. The first term in (19) is considered in order to penalize non-uniform vehicle coordination across the NOIR. The second term penalizes the non-uniform traffic inflow at the NOIR inlet road elements. Mathematically speaking, the above optimization problem can be formulated by the following quadratic programming problem:

$$\min\left(\frac{1}{2}\mathbf{U}^T\mathbf{\Gamma}_1\mathbf{U} + \mathbf{\Gamma}_2\right), \tag{20}$$

subject to inequality constraint

$$\mathbf{U} > \mathbf{0}, \tag{21}$$

and equality constraint

$$\|\mathbf{U}\|_1 = u_0\mathbf{1}_{n_\tau}, \tag{22}$$

where $\mathbf{1}_{n_t} \in \mathbb{R}^{n_\tau \times 1}$ is the one vector, $u_0 > 0$ is a constant scalar, and

$$\mathbf{\Gamma}_1 = \beta\mathbf{I}_{n_\tau N_{in}} + \mathbf{H}^T\mathbf{H}, \tag{a}$$

$$\mathbf{\Gamma}_2 = \mathbf{H}^T(\mathbf{G}\mathbf{X} + \mathbf{F}\mathbf{U}) \tag{b}$$

$$\mathbf{H} = \begin{bmatrix} \mathbf{0} & \mathbf{0} & \cdots & \mathbf{0} \\ \mathbf{W} & \mathbf{0} & \cdots & \mathbf{0} \\ \vdots & \vdots & \ddots & \vdots \\ \bar{\mathbf{Q}}_D^{n_\tau-2}\mathbf{W} & \bar{\mathbf{Q}}_D^{n_\tau-3}\mathbf{W} & \cdots & \mathbf{0} \end{bmatrix}, \tag{c}$$

$$\mathbf{F} = \begin{bmatrix} \mathbf{W} \\ \bar{\mathbf{Q}}_D\mathbf{W} \\ \vdots \\ \bar{\mathbf{Q}}_D^{n_\tau-1}\mathbf{W} \end{bmatrix}, \tag{d}$$

$$\mathbf{F} = \begin{bmatrix} \bar{\mathbf{Q}}_D \\ \bar{\mathbf{Q}}_D^2 \\ \vdots \\ \bar{\mathbf{Q}}_D^{n_\tau} \end{bmatrix}. \tag{e}$$

$$\tag{23}$$

## B. Microscopic Light-Based Responsive Control

The traffic responsive control is modeled as a receding horizon optimization (RHO) defined by the following fourtuple $(\mathbf{X}, \mathcal{T}, \mathcal{A}, \mathcal{J})$, where state vector $\mathbf{X}$ and transition function $\mathcal{T}(\mathbf{X}, \lambda) = \mathbf{Q}_\lambda \mathbf{X} + \mathbf{W}\mathbf{u}$ were previously defined in Section IV-B, $\mathcal{A}$ is a discrete action set, and $\mathcal{J}$ defines the RHO cost function. The movement cycle transition, defined by a nondeterministic finite state automaton, is fixed at every NOIR intersection $i \in \mathcal{V}_C$. However, the time interval of a movement phases can be optimally planned by the traffic lights.

The traffic actions are defined by set

$$\mathcal{A} = \{(a_{m_B+1}, \cdots, a_m) | a_i \in \{S, C\}, i \in \mathcal{V}_C\}, \quad (24)$$

where S and C are the abbreviations for "Switch" and "Continue", respectively, if $a_i = C$, the traffic light does not update the current movement phase at intersection $i \in \mathcal{V}_C$. Otherwise, $a_i = S$ selects the next movement phase which is unique and defined by a deterministic movement cycle graph (See the example movement cycle graph previously shown in Figure 2 (b)). A constant threshold time $T_{L,i}$ assigns an upper limit for the duration of command C at NOIR node $i$, i.e. C is overridden if the phase duration $\tau_{i,j}$ exceeds $T_{L,i}$ at NOIR node $i \in \mathcal{V}_C$ when $\lambda_{i,j}$ is active (See Figure 2 (b)).

The $N_\tau$-step expected cost is defined as follows:

$$\mathcal{J} = \sum_{\tau=1}^{N_\tau} (\mathbf{X}^T[k+\tau]\mathbf{\Gamma}\mathbf{X}[k+\tau]), \quad (25)$$

where $\mathbf{\Gamma} \in \mathbb{R}^{(N-N_{out}+N_{in}) \times (N-N_{out}+N_{in})}$ is a positive definite diagonal matrix. The matrix $\mathbf{\Gamma}$ can encapsulate interaction of pedestrians and environment with the traffic flow. In particular, the ii entry of $\mathbf{\Gamma}$ ($\Gamma_{ii} > 0$) is a scaling factor incorporating pedestrian tendency probabilities, as well as real-time traffic information including construction condition, road enclosures, traffic speed, special events, and incidents into planning.

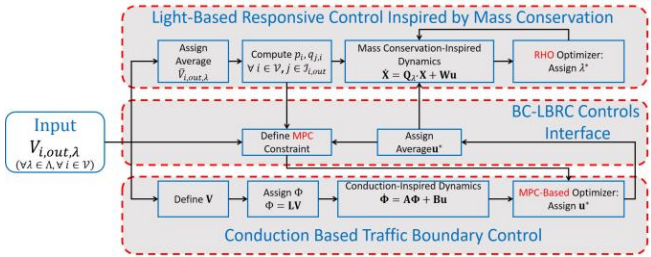

Figure 5: The functionality of the interactive BC-LBRC traffic control.

**Remark 4:** The functionality of the proposed IPBTC is shown in Figure 5 where an interface layer has been specified between the boundary and light-based controls. The interface layer has two main responsibilities. The first responsibility is to update the parameters of the conduction based dynamics using relation (16) given the probability matrix $\bar{\mathbf{Q}}$. Note that $\bar{\mathbf{Q}}$ is defined based on the tendency probabilities; $\bar{\mathbf{Q}}$ is continuously updated and incorporated into planning. The second responsibility is to specify the MPC constraints given in (20) and (21) based on the outflow traffic rates.

## VI. SIMULATION RESULTS

This section presents the simulation results for the example NOIR shown in Figure 6. The NOIR can be specified by GNOIR with boundary node set $\mathcal{V}_B = \{1, \cdots, 17\}$ and connection nodes defined by $\mathcal{V}_C = \{14, \cdots, 30\}$. The NOIR is heterogeneous consisting of unidirectional and bidirectional roads. This heterogeneous NOIR can be represented by a network with 53 unidirectional roads, where every unidirectional road is filled out by five serially-connected road elements. Elements in road $i$ are identified by the unique numbers $5(i-1) + j \in \mathcal{V}$ where $j = 1, \cdots, 5$. The BC traffic graph is defined by $\mathcal{G}(\mathcal{V}, \mathcal{E})$ where $\mathcal{V} = \{1, \cdots, 265\} = \mathcal{V}_{in} \cup \mathcal{V}_{out} \cup \mathcal{V}_I$, $\mathcal{V}_{in} = \{1, \cdots, 8\}$, $\mathcal{V}_{out} = \{45, 50, \cdots, 85\}$, and $\mathcal{V}_I = \mathcal{V} \setminus (\mathcal{V}_{in} \cup \mathcal{V}_{out})$.

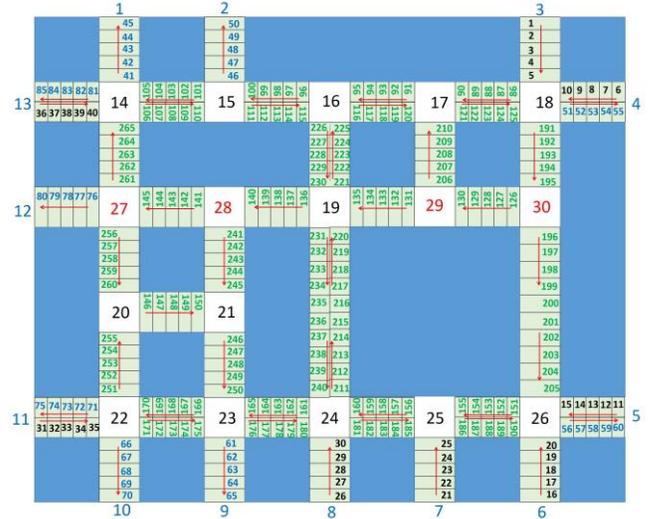

Figure 6: Example NOIR with 53 unidirectional roads. Every road is filled by five serially-connected road elements.

For the simulation, $u_0 = 54$ is selected. Therefore, the number of vehicles entering the NOIR is restricted to be 54 at any time. It is assumed that the number of movement phases at an intersection $i \in \mathcal{V}_C$ is the same as the number of incident roads to the intersection. Therefore, $|\lambda_{27}| = |\lambda_{28}| = |\lambda_{29}| = |\lambda_{30}| = 1, |\lambda_{15}| = |\lambda_{20}| = |\lambda_{21}| = |\lambda_{22}| = 2, |\lambda_{14}| = |\lambda_{16}| = |\lambda_{17}| = |\lambda_{18}| = |\lambda_{19}| = |\lambda_{23}| = |\lambda_{25}| = 3, |\lambda_{24}| = |\lambda_{26}| = 4$, and $|\Lambda| = |\lambda 1| \times \cdots |\lambda 30| = 559872$ is the total number of all movement phases defined by set $\Lambda$. It is assumed that there is no traffic light located at intersection node $i \in \{27, 28, 29, 30\} \subset \mathcal{V}_C$ as only one road is incident to these intersections.

Choosing $T_{L,i} = 3$, a road element cannot be active more than three time-steps, if it is incident to an NOIR connection node $i \in \mathcal{V}_C$ (Green light duration cannot exceed $T_{L,i} = 3$ time steps at every NOIR connection node). The optimal traffic light actions at junctions 14 through 26 are given in Figure 7 for $k = 1,\cdots,30$. Boundary control velocity inputs $\bar{V}_{1,in}$ through $\bar{V}_{8,in}$ are plotted versus discrete time $k$ for $k = 1,\cdots,30$ in Figure 8. Moreover, the potential values of all road elements are illustrated in Figure 9 at sample time $k = 50$. As shown in Figure 9, the potential value of every upstream node is greater than the potential value of its adjacent downstream element. Furthermore, $\phi_i[50] = 0$ for outlet road elements $45, 50, \cdots, 85$.

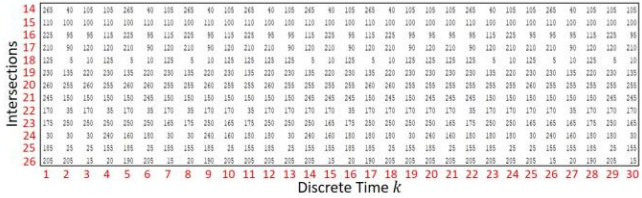

Figure 7: (a) Optimal movement phases at intersections 14 through 26 for $k = 1,\cdots,30$. The number shown in the plot specifies the active road element at time $k$.

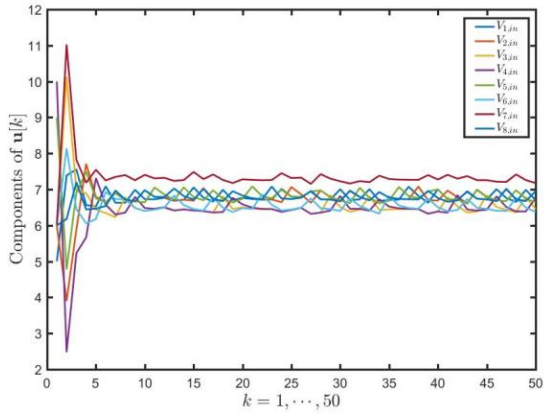

Figure 8: Boundary control input $\bar{V}_{1,in}[k]$ through $\bar{V}_{8,in}[k]$ for $k = 1,\cdots,30$.

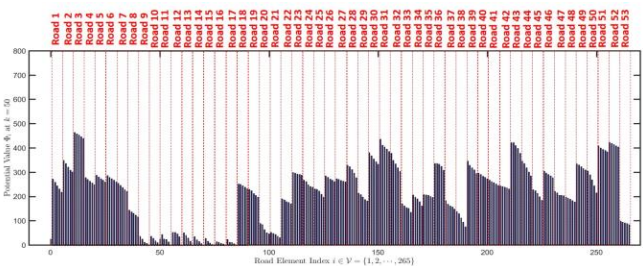

Figure 9: Potential values of all NOIR road elements at discrete time $k = 50$.

Figure 10 plots the potential value of the road element 162 versus time. It is seen that the average potential value stably decreases as $k$ increases.

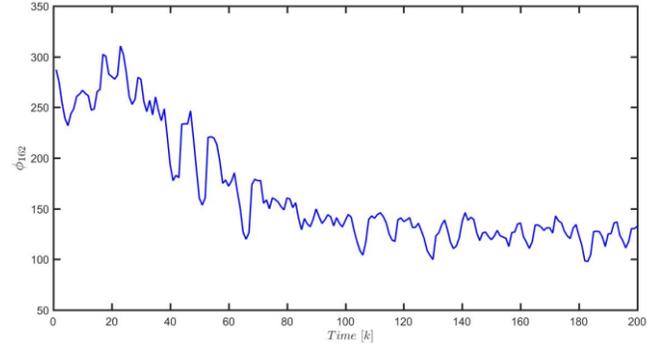

Figure 10: Traffic potential value of the road element 162 versus discrete time $k$ for $k = 1,\cdots,200$.

## VII. CONCLUSION

This paper develops physics-inspired approaches to model and control traffic congestion in a heterogeneous NOIR consisting of unidirectional and bidirectional roads. By developing an interactive physics-based traffic control, traffic congestion can be managed with minimal computation cost through controlling inflow rates at the NOIR boundary nodes as well as the NOIR intersections. This paper proposes a novel analogy between traffic coordination in a transportation system and the heat diffusion in a thermal system. This analogy offers several benefits that include but not limited to (i) consistent model-based learning of the macroscopic traffic parameters, (ii) developing a high-fidelity traffic coordination model, (iii) incorporating microscopic properties of a traffic system into planning, and (iv) enhancing the resilience of the traffic congestion control. The proposed IPBTC is computationally-efficient as it can successfully control the traffic congestion only by imposing constraints on the boundary control input, without imposing a constraint on the traffic state.

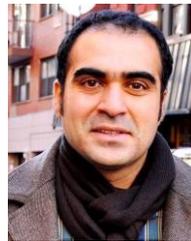

**Dr. Hossein Rastgoftar** is an Assistant Research Scientist in the Department of Aerospace Engineering at the University of Michigan. He received the B.Sc. degree in mechanical engineering-thermo-fluids from Shiraz University, Shiraz, Iran, the M.S. degrees in mechanical systems and solid mechanics from Shiraz University and the University of Central Florida, Orlando, FL, USA, and the Ph.D. degree in mechanical engineering from Drexel University, Philadelphia, in 2015. His current research interests include dynamics and control, multi-agent systems, cyber-physical systems, and optimization and Markov decision processes.